\documentstyle[aps,preprint,epsfig]{revtex}

\begin{document}
\preprint{TU-572}
\title{Axion model in extra dimensions with TeV scale gravity}
\author{Sanghyeon Chang, Shiro Tazawa and Masahiro Yamaguchi}
\address{Department of Physics, Tohoku University,
Sendai 980-8578, Japan}
\maketitle
\begin{abstract}
  A simple axion model is proposed in the scenario of large extra
  dimensions where gravity scale is as low as 1 TeV. To obtain an
  intermediate-scale decay constant of the axion, the axion is assumed
  to live in a sub-spacetime (brane) of the whole bulk. In this model
  there appear Kaluza-Klein modes of the axion which have stronger
  interaction than those of the graviton.  The axion brane plays
  a role of absorber of the graviton Kaluza-Klein modes. Because of
  these reasons phenomenology and cosmology of the axion become highly
  non-trivial. We discuss various cosmological constraints as well as
  astrophysical ones and show that the model is viable for certain
  choices of the dimensionality of the axion brane. The structure of
  the model proposed here provides a viable realization of the fat
  brane idea to relax otherwise very severe cosmological constraints.
\end{abstract} 

\section{Introduction}
It has been suggested that the fundamental scale of  nature can be
as low as TeV, whereas the largeness of the effective Planck scale or 
the weakness of the
gravity in a long distance can be explained by introducing large extra
dimensions\cite{ADD1,AADD,ADD2,ADM}.  When there exist $n$ of such extra 
dimensions, 
the relation between the gravitational constant $ 8\pi G_N=
1/M_{pl(4)}^2$ in $4$ dimensions and the fundamental scale $M_{pl(4+n)}$
in $(4+n)$ dimensions is given
\begin{equation}
M_{pl(4)}^2 \sim V_n M_{pl(4+n)}^{n+2} ,
\end{equation}
where $V_n$ is the volume of the extra dimensional space.
For $M_{pl(4+n)}\sim 1$ TeV, the size of the extra dimensions $r_n$ is 
computed as
\begin{equation}
r_n\sim V_n^{1/n}\sim 10^{\frac{32}{n}-6} M_{\rm TeV}^{-{2\over n} -1} 
{\rm MeV}^{-1}
\sim 10^{\frac{32}{n} -17} M_{\rm TeV}^{-\frac{2}{n} -1} {\rm cm} ,
\end{equation}
where $M_{\rm TeV} \equiv M_{pl(4+n)}/$TeV.
The case $n=1$ is excluded because the gravitational law would change at 
macroscopic level,  but the cases $n \geq 2$ are allowed by gravity 
experiments. 

There are also astrophysical and cosmological bounds \cite{ADD2}
because there are a tower of graviton Kaluza-Klein (KK) modes which contribute
to the 
supernova cooling \cite{hall}, 
the total mass of the universe and the late-time photon 
production \cite{photon} {\it etc.} 
We may avoid some of these problems by assuming an extremely low reheating
temperature of the early universe.

To have an inflation model with such a low temperature is already
a tough subject, especially with the cosmological moduli problem 
\cite{inflation}. 
Furthermore, we need baryon asymmetry after the
inflation while the temperature is quite low and we knew that
the process which generates baryon asymmetry should not induce proton
decay. It is better to have higher reheating temperature for many reasons,
but it is very hard to avoid known restrictions.

There is another problem in TeV scale gravity.
The simplest model of this type would not provide intermediate scales
which are necessary to explain phenomenological issues like, the
apparent gauge coupling unification \cite{DDG}, small neutrino masses
and the strong CP problem. It was pointed out \cite{ADD2} that
particles dwelling in the extra dimensions, other than graviton, can
have similar effective interaction terms in four-dimensional
physics.  This type of interaction is suppressed as gravity and thus
give extremely weak interaction between bulk matter and normal
matters.  For the neutrino, there have been many discussions and
researches about how to extend the minimal setting to obtain the small
neutrino masses and mixing \cite{neutrino,neutrino2,neutrino3}, but
the axion as a solution of the strong CP solution has not been studied
thoroughly in this context. In this paper we will propose an axion
model in TeV scale gravity for various numbers of the extra
dimensions, with special emphasis on cosmological constraints: the
addition of the axion in the large-extra-dimension model substantially
alters the cosmology.
 
In the following sections, we will use the convention
$M_{pl}\equiv M_{pl(4)}=$  reduced Planck mass and
$M_*\equiv M_{pl(4+n)}=$  fundamental scale.

\section{PQ scale in extra-dimension}

If an axion is a boundary field confined on
 4 dimensions, the Peccei-Quinn (PQ) scale $f_{PQ}$ is bounded by $M_* \sim
1$ TeV.  To obtain a higher PQ scale, the axion field has to be inevitably 
a bulk field. If it lives in the whole (4+n) dimensional bulk, the PQ
scale will be $f_{PQ} \sim M_{pl}$.

However, the damped coherent oscillation of the axion vacuum at
the early universe with $f_{PQ} \sim M_{pl}$ would overclose the universe. 
A conventional 
argument gives an upper bound $f_{PQ}\leq 10^{12}$ GeV. Even when an entropy
production takes place after the QCD phase transition ({\it e.g.} the
reheating temperature is smaller than $\sim 1$ GeV), $f_{PQ}$ cannot be 
much larger than $10^{15}$ GeV.

In Ref. \cite{kawa}, it was shown that the late thermal
inflation can raise bound of the PQ scale up to $10^{15}$ GeV. This
argument can be applied to our case, provided that the coherent
oscillation of the inflaton is followed by the reheating process. Then
the bound on $f_{PQ}$ is
\begin{equation}
f_{PQ} < 10^{15} \mbox{GeV} \left(\frac{h}{0.7}\right)
\left(\frac{\pi/2}{\theta}\right)
\left(\frac{\mbox{MeV}}{T_R}\right)^{1/2},
\end{equation}
where $h$ is from Hubble constant in units of 100km\,sec$^{-1}$Mpc$^{-1}$, 
$\theta$ is the initial value of PQ vacuum angle and $T_R$ is the reheating
temperature after inflation.
Thus the axion model suffers from the over-closure problem if there is
no other scale than $M_*$ and $M_{pl}$.

In this paper, we propose that a natural way to realize an intermediate
scale axion is to make it live in a $(4+m)$ dimensional sub-spacetime ($(3+m)$
brane) ($m<n$) of the whole $(4+n)$ dimensional bulk. The idea of using
sub-spacetime to realize an intermediate scale already appeared in Ref.
\cite{neutrino2} in the context of neutrinos.

Let $\chi$ be a complex scalar field which contains PQ axion in 4+m
dimension; $\tilde{a}$.
If the axion field lives only on $4+m$ dimensional sub-spacetime 
where $m<n$ and  the volume of extra-dimension is $V_m$,
\begin{eqnarray}
{\cal L}_\chi =   \int dx^{4+m} \partial^M \chi^* \partial_M \chi
+ \int dx^4 \frac{\tilde{a}(x^A=0)}{\langle \chi\rangle} F\tilde{F}  
\end{eqnarray}
where $x^A$ means an extra-dimension coordinate. Assuming that the
vacuum expectation value of the $\chi$ field does not depend on the
extra-dimension coordinates, we obtain 
\begin{eqnarray}
f_{PQ}&\sim& \sqrt{V_m}\langle \chi \rangle \sim r_n^{m/2} M_*^{1+m/2}
\sim M_*\left(\frac{M_{pl}}{M_*}\right)^{m/n}
\nonumber \\
&\sim& 10^{3(1+5m/n)} M_{\rm TeV}^{1-m/n} \mbox{GeV} .
\end{eqnarray}
Here we have defined the 4-D axion field as $a= \sqrt{V_m}\tilde{a}(x^A=0)$ 
and assumed that the size $r_n$ is common for all extra dimensions.

A lower bound of $f_{PQ}$ comes from astrophysical observations, e.g.
red giant and supernova cooling by axion emission.
It is known that $f_{PQ}$ should be larger than $10^{9}$GeV.
In extra dimension physics, KK modes also contribute to supernova cooling
if their masses are smaller than the core temperature ($\sim 30$MeV).
We discuss it in the next section.

To have $10^{9}$ GeV $<f_{PQ} \leq 10^{15}$ GeV, we need $2/5<m/n \leq
4/5$.  Possible sets of ($m,n$) with $f_{PQ}$ where $M_*= 1$ TeV and
$10$ TeV can be found in Table \ref{table1}, \ref{table2}.

\section{Laboratory and Astrophysical constraints}
Experiments on detecting axion from the nuclear reactor and the sun give
bounds on PQ scale $<10^4$ GeV (Lab), $10^4$ GeV $< f_{PQ} <10^6$ GeV from the sun.
However this limit strongly depends on the photon axion coupling and 
method of detecting axion.
Furthermore the laboratory bound is much weaker than the astrophysical bound.

The strongest bound from astrophysics is the Supernova cooling.
In SN1987A observation, it was calculated that 
\begin{equation}
\frac{1}{f_{PQ}^2} < 10^{-18} \mbox{ GeV}^{-2}.
\end{equation}
Since the axion KK modes interact exactly the same way as the conventional 
axion, 
the effective interaction of the KK modes at the
core temperature $T\simeq 30$ MeV is
\begin{equation}
\frac{1}{f_{PQ}^2}\times (T r_n)^m \sim \frac{T^m}{M_*^{m+2}} <10^{-18}
\mbox{ GeV}^{-2}.
\end{equation}
This gives a bound on the fundamental scale
\begin{equation}
M_* > (10^{18} \times 0.03^m)^\frac{1}{m+2} \mbox{ GeV}.
\end{equation}
For $m=1$, $M_*>300$ TeV
\footnote{It was argued that there arise large quantum corrections
in the case of one extra dimension in general, which would destabilize
the gauge hierarchy \cite{AB}. This argument would exclude the $m=1$ case.}
 and $m=2$, $M_* > 5$ TeV. $M_* \sim 1$ TeV is allowed
only if $m>2$.
This suggests that the number of extra-dimension should be at least $3$ to
include PQ mechanism as a solution of strong CP problem in TeV scale gravity
model.

In near future, high energy accelerator experiments may probe 
the axion KK mode emission.   The graviton KK mode signals in collider
were discussed in detail recently. \cite{han2}
The scattering cross section of the graviton KK mode emission from high energy 
scattering with center of mass frame energy $\sqrt{s}$ is 
\begin{equation}
\sigma \propto \frac{1}{M_{pl}^2} (\sqrt{s}r_n)^n
\sim \left(\frac{\sqrt{s}}{M_*}\right)^{n+2}  \frac{1}{s},
\end{equation}
while the KK axion production is
\begin{equation}
\sigma \propto \frac{1}{f_{PQ}^2} (\sqrt{s}r_n)^m
\sim \left(\frac{\sqrt{s}}{M_*}\right)^{m+2} \frac{1}{s}.
\end{equation}
Since the energy dependence of the axion KK mode
cross-section is different from the
graviton KK mode cross-section,
it might be possible to detect 
this difference at TeV scale collider experiments. 

\section{Thermal production of Axion KK mode}
Since the axion dwells in the extra-dimensional brane,
their masses are proportional to $r_n^{-1}$ and
they have stronger couplings than the graviton KK modes to the normal 
matters in general cases.

If there is no hidden particle which couples to the axion or the mass
of the  KK mode axion is lower than the sum of three pion masses $\sim 500$ MeV,
the main decay channel of light  KK axion is to two photons.
The life-time of each KK mode for a given $(n,m)$ can be found in Table
\ref{table1}, \ref{table2}.
\begin{equation}
 \Gamma_{a_{KK}\rightarrow 2\gamma} \simeq \frac{C_{a\gamma}^2}{64\pi} 
\left(\frac{\alpha}{\pi}\right)^2 \frac{m_{A}^3}
{f_{PQ}^2} \sim 3\cdot10^{-8} C_{a\gamma}^2  \frac{m_{A}^3}{f_{PQ}^2},
\end{equation}
where $m_A$ is the mass of the axion KK mode and
$C_{a\gamma}$ is the model dependent axion-photon coupling
which is usually within $0.1$ to $1$.
This decay can be  cosmologically dangerous.
For instance, for $f_{PQ}= 10^{12}$ GeV and $m_{A}=1$ MeV,
 life time of KK mode is $\tau_{A} \sim 10^{17}$ sec, which is about the age of the universe.

The graviton KK modes have similar cosmological problems
because they can overclose our universe or decay into the photons at a
late stage of cosmological evolution.  Originally it was suggested
that a ``fat brane" \cite{ADD2} can solve the cosmological problems by
absorbing most of the decay products of the KK modes.  However
massless particles in the higher dimensional brane {\it are not}
massless in our four-dimensional universe since they have momenta in
the extra dimensions which appear masses in our universe \cite{hall}.
Another way to avoid this difficulty is assuming a large number of
four-dimensional lattices in the bulk. But we need at least $10^6$
{\it empty} universes. Or we should assume the existence of a
four-dimensional hidden sector which has $10^6$ times more degrees of
freedom than those of the standard model  while they should not be
produced significantly by the reheating process after the inflation.

If we add an axion as a ``brane particle" in this model, the graviton
``bulk particle" will decay to the ``brane" axion more efficiently, 
since its decay width will be enhanced by factor $(M_G r_n)^m$.
The graviton KK mode with mass $m_G$ will have decay width to the axion
\begin{equation}
\Gamma(g_{KK} \rightarrow 2a) \sim {\cal O}(10^{-3})\frac{m_G^3}{M_{pl}^2}
\times (m_G r_n)^m.
\end{equation}

After some period, 
instead of the massive graviton KK fields, we will have the axion 
KK fields with about same masses.
Since the massive axion KK mode can decay into the photon pairs 
and the life time
of the graviton KK mode becomes much shorter, the primordial graviton KK mode
will not over-close the universe. Instead, it will contribute 
to the cosmological
background radiation. This can be a severe constraint to the axion model. 
Also axions can be produced thermally during the reheating process. 
For these reasons, we have to check whether the axion model can survive 
the cosmological constraints.

To estimate the constraints for various cases, we calculate the amount
of the thermal axion produced at the reheating temperature
 $T_R$ and the axion KK modes from the graviton KK modes decay.
In  Appendix A, we derived the Boltzmann Eq. for the yield $Y= \rho/s$ 
for four different sources of the axion KK modes:\\ 
{\bf I.} decay of 
graviton KK modes from the inverse decay, ($2\gamma , e^+e^-, \bar{\nu}\nu
\rightarrow g_{KK}$); the energy of the KK modes are 
concentrated on $m_{A}\sim T_R$,
\\ {\bf II.} decay of  graviton KK modes produced from the scattering, ($e\gamma \rightarrow e 
g_{KK},
 e^+e^-\rightarrow\gamma g_{KK}$); this contribution is significant only if
$m_{A} < T_R$,\\
{\bf III.} the axion KK mode from the pion scattering ($\pi\pi \rightarrow \pi
a_{KK}$); this process dominates if $T_R > 10$ MeV \\
{\bf IV.} the axion KK mode from the two photon inverse decay  ($2\gamma \rightarrow
a_{KK}$); this gives significant contribution when $m_{A}\sim T_R$.
 
For each case, 
\begin{eqnarray}
Y_1 &\simeq&  3\cdot 10^{-23} \left(\frac{T_R}{100\mbox{MeV}}\right) A_1,\\
Y_2 &\simeq&  2\cdot 10^{-23} \left(\frac{T_R}{100\mbox{MeV}}\right) A_2,\\
Y_3 &\simeq& 6\cdot 10^{-10} \left(\frac{10^{12}\rm GeV}{f_{PQ}}\right)^2
\left(\frac{T_R}{100\mbox{MeV}}\right)^3 A_3,\\
Y_4 &\simeq& 2\cdot 10^{-16} \left(\frac{10^{12}\rm GeV}{f_{PQ}}\right)^2 
\left(\frac{ T_R}{100\rm MeV}\right) A_4
\end{eqnarray}
where
\begin{eqnarray}
A_1&=& \left(\frac{10}{g_*(T_R)} \right)^{3/2}
\left(\frac{m_{A}}{T_R} \right)^3 \\
A_2&=& \left(\frac{10}{g_*(T_R)} \right)^{3/2}
\left(\ln\left(\frac{T_R^3}{m_{A}^2m_e}\right)-0.8 \right)\\
A_3&=& C_{a\pi}^2\left(\frac{10}{g_*(T_R)} \right)^{3/2} 
\left(\frac{I(T_R)}{1000}\right)\\
A_4&=& C_{a\gamma}^2\left(\frac{10}{g_*(T_R)} \right)^{3/2}
\left(\frac{m_{A}}{T_R} \right)^3.
\end{eqnarray}
For the details of these calculations and the definition of
function $I(T)$, see appendix  A (we present the numerical
plot of $I(T)$ in Fig.~5). 

\section{Cosmological constraints}
In this section we would like to discuss various cosmological constraints on
the model for a given $(m,n)$ set.
Notice that in Table I and II, $n\geq 5$ in both $M_*=1 $ and
$10$ TeV case and $n=4$ in $1$ TeV are 
cosmologically safe if $T_R$ is low enough ($\sim 1$ MeV).
If the minimal KK mode mass is greater than $1$ MeV, 
the KK modes are not generated in the thermal bath of such a low reheating
temperature.
On the other hand $m \leq 2$ in $M_*=1 $ TeV and $m=1$ in $M_*=10$ TeV
is forbidden by the astrophysical bound.
The cases $n=4, m=3$ at $M_*=1$ TeV and  $n=3, m=2$ at $M_*=10$ TeV are
not trivially allowed or ruled out by the cosmological constraints.
Details on these cases are discussed in Appendix B.

\subsection{Big bang nucleosynthesis}
At the temperature of the universe around $1$ MeV, it is required that
there should not be additional particles which contribute to
the energy density significantly. Otherwise $^4$He would be produced
more than what is observed now because the universe should expand faster than
the standard scenario.

We apply a rather loose bound that the energy density contribution
by the KK mode should be smaller than  one
neutrino energy density at $T =1$ MeV.
At high $T_R(>10$ MeV$)$, axion KK mode production dominated by process III. 
In case III, one may practically have maximal mass 
\begin{equation}
m_1\equiv \max\{m_\pi, T_R\} \label{eq:m1}
\end{equation}
for the KK mode which is produced in the thermal bath.

Thus the bound from  BBN is
\begin{eqnarray}
\left.\frac{\rho_A}{s}\right|_{BBN} &\simeq &
D \int^{m_1}_{m_0} dm_A \times (m_Ar_n)^m Y_A 
\nonumber\\
&\simeq& m_1 Y_3 (m_1 r_n)^m
\simeq
2 \cdot 10^{-3} \frac{M_{pl} T_R^3m_1^{m+1}}{M_*^{m+2} f_\pi^2}
A_3  < 0.1 \mbox{ MeV} ,
\end{eqnarray}
where $D$ is normalization constant which is equal to
$m$ in torus compactification
with universal distance $r_n$. In our calculations we restrict ourselves
to this case.  

If $T_R\sim m_\pi$, approximately 
\begin{equation}
M_*>10^{\frac{20}{m+2}} \times m_\pi.
\end{equation}
For  $m=1$, this reads $M_* > 600$ TeV, and for $m=2$, $M_* >15 $ TeV.
But if the reheating temperature is as low as $10$ MeV, this bound 
is not important since $A_3(T_R)$ is suppressed exponentially.

\subsection{Over-closure of Universe}
The total energy of the axion KK modes at present must not exceed the
critical density:
\begin{equation}
\rho_A <\rho_c = 3\cdot10^{-6} s_0 h^2 \mbox{ MeV} ,
\end{equation}
where $s_0\simeq 3000$cm$^{-3}$ is the entropy of the present universe.

For the case that the KK modes decay into some relativistic particles,
we can divide the bound in two parts; decay before the present
time and do not decay till now:
\begin{equation}
\frac{\rho_A}{s_0}\simeq D\int^{m_2}_{m_0} dm_A (m_A r_n)^m Y_A
+D\int^{m_1}_{m_2} dm_A (m_A r_n)^m \frac{Y_A T_0}{T(m_A)}
< 3\cdot 10^{-6} h^2 \,{\rm MeV},
\end{equation}
where the axion KK mode with mass $m_2$ decays at the present time, and $m_1$
is defined in Eq.~(\ref{eq:m1}).

\subsection{Light element destruction}
Energetic photons from  heavy axion KK modes which decay after 
$10^4$ sec can destroy the light elements made during the nucleosynthesis.
Therefore there are several bounds on the density of the KK modes which
weigh than  $10$ MeV. They are \cite{ellisetal}
\begin{equation}
\frac{\rho_{A}}{s}\leq 10^{-12} \mbox{GeV}
\end{equation}
for $\tau_{A} \geq 10^7$ sec,
\begin{equation}
\frac{\rho_{A}}{s}\leq  10^{-6} - 10^{-10} \mbox{GeV}
\end{equation}
for $10^4 <\tau_{A} \leq 10^7$ sec.
Usually these bounds are less important than cosmological microwave 
background bound given below.

\subsection{Cosmological Microwave Background Radiation}
If the massive KK modes decay after 
$\tau_{A} \geq 10^6$ sec but before the recombination era, the produced
photons may give a distortion of the cosmological microwave background
radiation. The COBE observation gives a bound \cite{CMBR}
\begin{equation}
\frac{\Delta \rho_\gamma}{s} \leq 2.5 \times 10^{-5} T_D
\end{equation}
where $T_D$ is temperature at KK mode decay.

\subsection{Diffuse photon background}
Observations of diffuse photon backgrounds at the present universe give
upper bounds on additional contributions to photon spectrum. For example,
for the energy range $800$ keV $<E <30$ MeV \cite{diffuse-photon}
\begin{equation}
\frac{d{\cal F}}{d\Omega} \simeq E \times A \left(E/E_0\right)^{-\alpha}
\simeq 78 \left(\frac{E}{ 1 {\rm  keV}}\right)^{-1.4}.
\end{equation}
Constraints on other ranges of the photon energy can be found, e.g. in 
Ref.~\cite{KY}

Theoretical prediction is
\begin{equation}
\frac{d{\cal F}}{d\Omega} = \frac {n_{A} c}{4\pi} \times Br
\end{equation}
for the life-time of KK mode shorter than the age of the universe, and
\begin{equation}
\frac{d{\cal F}}{d\Omega} \sim Br \times\frac {n_{A} c}{4\pi}
\frac{\Gamma_{a_{KK}\rightarrow
2\gamma} }{Br H_0} \left( \frac{2E}{m_Ac^2}\right)^{3/2} (m_Ar_n)^m
\end{equation}
for its life time  longer than the age of the universe \cite{Kolb-Turner}.
Here we have introduced $Br$ as a branching ratio of axion decay into two photons.

In our brane picture, there is {\it a priori} no reason that
the axion brane contains only our four-dimensional wall. Rather it
will be natural to imagine that there is a parallel universe(s) or
another four-dimensional wall in the brane. Or one can just imagine that
there are some unknown particles on our wall itself.  Then one can consider the
situation that thermally produced KK modes of graviton will mainly
decay into the axion in the brane and axions both from the graviton
decay and the thermal production will decay into the parallel wall if
this wall has some kind of QCD and/or $U(1)$ type of interactions.
(Or it can decays into some hidden QCD/$U(1)$ fields in our universe.)
Since the axion decay width is highly suppressed
by $\left(\frac{\alpha}{\pi}\right)^2$ with $\alpha$ being the fine structure
constant,
it is easy to get a low branching ratio to decay into the photon, in
other words, most of the axions decay into the other wall(invisible section)
if the
coupling constant, the color factor of the other gauge interaction,
and/or the number of the fermions with PQ charges in the decay loop
diagram in the other wall(invisible section) are large enough.

\subsection{Results}

Here we summarize the
results we obtained. The reader should be referred to Appendix B for more
detail. (We approximate $h=0.7$)\\
{\bf I.  $n=4,\ m=3$ and $M_*=1$ TeV case:} \\
1. BBN bound 
\begin{equation}
T_R <90\, A_3^{-1/3} \mbox{ MeV} \simeq 80  \mbox{ MeV}.
\end{equation}
2. Over-closure bound,
\begin{equation}
T_R <  12\, 
\left(\frac{C_{a\gamma}^2}{A_3^2 Br}\right)^{\frac{1}{6}} 
\mbox{ MeV}
\simeq 30 \mbox{ MeV  (for } Br=1) .
\end{equation}
3. CMBR bound, (for $m_A\geq 100$ MeV and $T_R>10$ MeV)
\begin{equation}
T_R < 2 \cdot 10^{-2} 
\left(\frac{C_{a\gamma}^2}{A_3^2 Br^3}\right)^{\frac{1}{6}} 
\mbox{ MeV} .
\end{equation}
4. Diffused photon bound, \\
for $T_R> 10$ MeV, 
\begin{equation}
T_R< 2\cdot 10^{-2} \left(\frac{Br}{C_{a\gamma}}\right)^{-0.63} \left(
\frac{m_A}{10\rm MeV}\right)^{-0.53}A_3^{-1/3} \mbox{ MeV},
\end{equation}
for $T_R< 10$ MeV, 
\begin{equation}
T_R< 0.3\, Br^{-0.73} \mbox{ MeV},
\end{equation}
where $Br<\Gamma_{a\rightarrow 2\gamma}/H_0$, and 
\begin{equation}
T_R < 5\, C_{a\gamma}^{-0.48} \mbox{ MeV}
\end{equation}
where $Br > \Gamma_{a\rightarrow 2\gamma}/H_0$.

\noindent{\bf II. $n=3,\ m=2$ and $M_*=10$ TeV case:} \\
1. BBN bound 
\begin{equation}
T_R < 100\, A_3^{-1/3} \mbox{ MeV} \simeq 90 \mbox{ MeV}.
\end{equation}
2. Over-closure bound,
\begin{equation}
T_R <  28\, 
\left(\frac{C_{a\gamma}^2}{A_3^2 Br}\right)^{\frac{1}{6}} \mbox{ MeV} \simeq
40 \mbox{ MeV (for } Br=1).
\end{equation}
3. CMBR bound,
\begin{equation}
T_R < 4 \cdot 10^{-2} 
\left(\frac{C_{a\gamma}^2}{A_3^2 Br^3}\right)^{\frac{1}{6}} 
\mbox{ MeV}.
\end{equation}
4. Diffused photon bound,\\
for $T_R> 10$ MeV, 
\begin{equation}
T_R< 3\cdot 10^{-2} \left(\frac{Br}{C_{a\gamma}}\right)^{-0.63} \left(
\frac{m_A}{10\rm MeV}\right)^{-0.2}A_3^{-1/3} \mbox{ MeV},
\end{equation}
for $T_R< 10$ MeV, 
\begin{equation}
T_R< 0.1\, Br^{-1.2} \mbox{ MeV},
\end{equation}
where $Br<\Gamma_{a\rightarrow 2\gamma}/H_0$, and 
\begin{equation}
T_R < 3 \mbox{ MeV}
\end{equation}
where $Br > \Gamma_{a\rightarrow 2\gamma}/H_0$.

In the both cases, $n=4,\ m=3$ and $M_*=1$ TeV and $n=3,\ m=2$ and
$M_*=10$ TeV, life times of axion KK modes up to $m_{A}\sim 1$ GeV are
quite long and so the constraint from the light element destruction
will not give any further bound.

We also performed computer calculations on both $n=4,\ m=3$ and
$M_*=1$ TeV
and $n=3,\ m=2$ and $M_*=10$ TeV cases. Fig. 1 and 2 show that
the DPBR is the stringent bound if the branching ratio is small. But in an
extremely small
branching ratio case, CMBR is dominant. This is because the life time of the
axion KK modes becomes shorter than $10^{13}$ sec, and so the produced photons
will disturb the CMBR spectrum.  One can see that BBN is independent
of the branching ratio. Note that these behaviors are consistent with
what we observed from the analytical computations given above.

In Fig. 3 and 4, we present combined cosmological limits for
all possible combinations of $n, m$ in Table I and II.  We find that the
reheating temperature is allowed to be significantly large if the branching
ratio to the photon is small enough.

\section{conclusions}

In this paper, we discussed the axion model in the  extra dimensions whose 
PQ scale lies in an
intermediate scale $f_{PQ} \leq 10^{15}$ GeV. This intermediate scale can
be obtained by introducing a $3+m$ dimensional brane in the 
$4+n$ dimension bulk.

If we include the axion as a brane particle, it will change the phenomenology
of the extra dimension physics, especially, cosmology.
Since the graviton KK mode will decay into the axion KK mode, the
over-closure problem is not as serious as the original model of Arkani-Hamed
et.al. 
On the other hand, the argument from stars and supernova cooling will
give a more strict bound on the axion production. 
Among other things, the most severe cosmological bound comes from photon 
emission through the decays of the KK modes of the axion. 
We found that the astrophysical argument restrict the number of the dimensionality of the sub-spacetime where the axion lives: $m>2$ for $M_*=1$ TeV and 
$m>1$ for $M_*=10$ TeV.
The latter cosmological argument requires quite a low reheating temperature
after the inflation.

To lift this bound, we can introduce the hidden matter/gauge fields to another four-dimensional wall (or even to our
wall itself) which has much stronger coupling to
axion and/or much more generations of particles, (or maybe much lower
QCD phase transition scale, etc). This can significantly lower the
branch ratio of the axion KK mode decay into photons.

The whole picture can be used to improve the original {\em fat brane} idea. 
This higher dimensional object plays a role of an absorber of the KK graviton
modes. If the fat brane couples to four dimensional wall(s) with interaction stronger than gravity, the produced particles in the fat brane may then decay
into relativistic particles on the four dimensional wall(s). This mechanism
can solve the problem of 
the overclosure of the universe by the KK modes.  Note that the fat-brane 
particles 
are not necessarily the axions as we discussed, but can be any  other weakly
interacting particles living in a higher dimensional brane. Moreover if 
the  particles produced by the fat-brane particle decays do not contain 
photons or any other cosmologically dangerous
particles, then we can avoid other cosmological problems such as the ones related to the cosmic photon backgrounds. It is worth mentioning that 
some mechanisms of generating small neutrino masses have a similar structure,
which will be discussed elsewhere \cite{CTY2}

In this paper, we did not consider an alternative solution to the
strong CP problem such as a spontaneous CP breaking model. This
typically requires very heavy quarks. If we want to keep all particles
carrying standard-model gauge charges in four dimensions, the maximum
scale we can have is the fundamental scale i.e. around TeV. For this
reason, this type of model is not favorable in the extra-dimension
scheme.  

\acknowledgments
This work was supported in part by the Grant-in-Aid for Scientific 
Research from the Ministry of Education, Science, Sports, 
and Culture of Japan, 
on Priority Area 707 "Supersymmetry and Unified Theory of Elementary
Particles", and by the Grant-in-Aid No.11640246 and  No.98270.
SC thanks the Japan Society for the Promotion of Science for financial support.

\appendix
\section{Estimation of the primordial axion KK mode density}
The production rate of the axion KK mode from initial particle $i$ and $j$
can be calculated from Boltzmann Eq.,
\begin{equation}
\dot{n}_{A} + 3 H(t) n_{A} = \sum_{ij} \langle \sigma v\rangle_{ij} n_i n_j,
\label{boltz1}
\end{equation}
where $n_A, H(t), \sigma, n_i$ represent axion KK mode number density,
Hubble constant, scattering cross section and number density of initial
particle $i$.

Using the relation $t=0.5 H(T)^{-1}$, we can convert the
time parameter to inverse temperature $x\equiv m/T$,
where
$H(T) \simeq \left(\frac{g_*}{10} \right)^{1/2}
\frac{T^2}{M_{pl}}$.
If we assume that the KK modes are produced by the particles in equilibrium,
we can rewrite (\ref{boltz1}) with the yield $Y\equiv n_{A}/s$. 
($s$ is entropy density
$s=\frac{2 \pi^2}{45} g_{*s} T^3$,
where $g_{*s}\simeq g_*$ is approximately 10 for 1 MeV $< T <$ 100 MeV.)
\begin{equation}
\frac{dY}{dx}=\frac{x}{H(m)}\Gamma_{A} Y_{EQ} \label{scat1}
\end{equation}
where 
\begin{equation}
\Gamma_A = n_{EQ}\sum_{ij} \langle \sigma v\rangle_{ij} .
\end{equation} 
A similar equation can be derived for the inverse decay case.
\begin{equation}
\frac{dY}{dx}=\frac{x}{H(m)} \Gamma\left(a_{KK}\rightarrow
\mbox{ALL}\right)\left\langle\frac{m_{A}}{E_{A}}\right\rangle Y_{EQ} 
\label{decay1}
\end{equation}
The yield of the KK mode at the equilibrium $Y_{EQ}$ is about 
$0.28/g_{*s}$ when initial particles are
relativistic (or proportional to $\exp(-x)$ if they are non-relativistic).
After integrating  Eq.(\ref{scat1}) and (\ref{decay1}) 
from the reheating temperature
$T_R$ to present temperature, we will get result with form 
\begin{equation}
Y \simeq \frac{\Gamma}{H(T_R)} Y_{EQ}(T_R)\sim 3
\cdot 10^{-2} \left(\frac{10}{g_*(T_R)}
\right)^{3/2}\frac{M_{pl}\Gamma }{T_R^2},
\end{equation} 
which can be used most of calculations reliably.

Let's estimate the sources of the axion KK mode production. 
The KK modes of 
axion can be produced from either
the thermal graviton KK mode decay or initial thermal
bath. 
We will classify four relevant cases.

\noindent{\bf Class I}: The KK mode of graviton which has mass around 
the reheating temperature 
$m_{G}\simeq T_R$ generated dominantly through the inverse decay $\gamma\gamma \rightarrow
g_{KK}$,  $\bar{\nu}\nu  \rightarrow g_{KK}$ and
 $e^+ e^-  \rightarrow g_{KK}$ \cite{han2}
\begin{eqnarray}
\Gamma_{g_{KK}\rightarrow 2\gamma}& =& \frac{m_{G}^3}{80\pi M_{pl}^2}\\
\Gamma_{g_{KK}\rightarrow f\bar{f}}& =& \frac{m_{G}^3}{160\pi M_{pl}^2}
\end{eqnarray}
for initial spin averaged.
For the tensor mode of graviton KK mode we should multiply 5 to $\Gamma$.
This will generate
\begin{equation}
Y_1 \simeq 6 \times 10^{-4} \frac{T_R}{M_{pl}} A_1
\simeq 3\cdot 10^{-23} \left(\frac{T_R}{100\mbox{MeV}}\right) A_1.
\label{Y1}
\end{equation}
where 
\begin{equation}
A_1= \left(\frac{10}{g_*(T_R)} \right)^{3/2}
\left(\frac{m_{A}}{T_R} \right)^3.
\end{equation}
Here we approximated that $m_A\simeq m_G$ after the graviton decay.
This mode is most abundant at $m_{A}=T_R$ and decrease fast if
$m_{A} \ll T_R$.

\noindent{\bf Class II}:
The KK modes which has much less mass than $m_{G}<T_R$ will be
produced dominantly by the scattering processes 
$e^{\pm}\gamma \rightarrow g_{KK} e^{\pm}$,
$e^+ e^- \rightarrow g_{KK} \gamma$.
If we choose a limit that KK mode mass is less than $T_R$ but
greater than the electron mass, we can calculate the interaction rate
from the amplitude presented in \cite{han2}.
\begin{eqnarray}
\Gamma_{e^{\pm}\gamma\rightarrow g_{KK}e^{\pm}} &\simeq& \langle \sigma v\rangle
n_{EQ} \simeq \frac{\alpha}{M_{pl}^2}\left(\ln\frac{T_R^3}{m_{G}^2m_e} 
-\frac{7}{8}\right) \times (0.3) T_R^3 \\
\Gamma_{e^+e^-\rightarrow g_{KK}\gamma} &\simeq&
\frac{\alpha}{6 M_{pl}^2} \times (0.3) T_R^3
\end{eqnarray}
The yield from scattering is
\begin{equation}
Y_2 \simeq 2\cdot  10^{-23} \left(\frac{T_R}{100\mbox{MeV}}\right)
A_2,
\label{Y2}
\end{equation}
where 
\begin{equation}
A_2= \left(\frac{10}{g_*(T_R)} \right)^{3/2}
\left(\ln\left(\frac{T_R^3}{m_{A}^2m_e}\right)-0.8 \right).
\end{equation}
for each KK mode with mass $m_{A}$.
This bound is valid only if $A_2$ is positive, i.e. the reheating
temperature is significantly higher than the KK mode mass.

\noindent{\bf Class III}: Thermal axion produced mainly by pion-pion scattering 
$\pi^\pm \pi^0 \rightarrow \pi^\pm a,$ $\pi^+\pi^- \rightarrow \pi^0 a,$
for 
$T_R > 10$ MeV. 
\begin{equation}
\Gamma_{2\pi\rightarrow a\pi} Y_{EQ} =\frac{3}{1024 \pi^5} 
\frac{C_{a\pi}^2}{f_{PQ}^2 f_\pi^2} T^5 
\left(\frac{45}{2\pi^2 g_{*s}}\right)
\times I(T) 
\end{equation}
where $C_{a\pi}$ is $(1-z)/3(1+z)$ with $z= m_u/m_d$. In the limit of
$m_A=0$, 
we can used the temperature dependent function $I(T)$ in Ref.\cite{chang} 
\begin{equation}
I(T)\equiv \int dx_1 dx_2\frac{x_1^2 x_2^2}{y_1 y_2}f(y_1)f(y_2)\int^1_{-1} d\omega
\frac{(s-m_\pi^2)^3(5s-2m_\pi^2)}{s^2 T^4}
\end{equation}
where $f(y)= 1/(e^y-1)$, $x_i=|\vec{p_i}|/T$, $y_i = E_i/T$ $(i=1,2)$
and $s = 2( m_\pi^2 + T^2(y_1 y_2 - x_1 x_2 \omega) )$.
We justify ourselves using $m_{A}=0$ limit by presenting
 the plot produced by computer
(Fig. 5) which shows that the mass dependence of $I(T)$ is indeed small.

$I(T)$ is around $10^3$ for $T>50$ MeV, and it suppressed exponentially
at $T<10$ MeV. 
For $T\ll m_{\pi}$, we can approximate function $I(T)$,
\begin{equation}
I(T) = \frac{\pi}{8}\left(\frac{3m_\pi}{T}\right)^5 \exp\left(-2\frac{m_\pi}{T}
\right).
\end{equation}
We can estimate thermal axion KK mode yield from pion scattering,
\begin{equation}
Y_3 \simeq 2\cdot
10^{-3} \frac{M_{pl}
T_R^3}{f_{PQ}^2 f_\pi^2} A_3
\simeq 6\cdot 10^{-10} \left(\frac{T_R}{100\mbox{MeV}}\right)^3
\left(\frac{10^{12}\rm GeV}{f_{PQ}}\right)^2 A_3
\label{Y3}
\end{equation}
where 
\begin{equation}
A_3= \left(\frac{10}{g_*(T_R)} \right)^{3/2} C_{a\pi}^2
\left(\frac{I(T_R)}{1000}\right).
\end{equation}

\noindent{\bf Class IV}: Thermal axion can be generated through the two photon 
inverse decay,
\begin{equation}
 \Gamma_{a_{KK}\rightarrow 2\gamma}
 \simeq \frac{C_{a\gamma}^2}{64\pi} \left(\frac{\alpha}{\pi}\right)^2
\frac{m_{A}^{3}}
{f_{PQ}^2} \simeq 2.7\cdot 10^{-8} C_{a\gamma}^2  \frac{m_{A}^{3}}{f_{PQ}^2},
\end{equation}
which leads 
\begin{equation}
Y_4 \simeq 8\cdot 10^{-10} \frac{M_{pl} T_R}{f_{PQ}^2 } A_4
\simeq 2\cdot 10^{-16} \left(\frac{10^{12}\rm GeV}{f_{PQ}}\right)^2 
\left(\frac{ T_R}{100\rm MeV}\right) A_4
\label{Y4}
\end{equation}
where
\begin{equation}
A_4= C_{a\gamma}^2\left(\frac{10}{g_*(T_R)} \right)^{3/2}
\left(\frac{m_{A}}{T_R} \right)^3.
\end{equation}
This shows that the axion produced from the inverse decay will dominate
over the axion from the pion scattering {\bf III}, if $T_R < 10$ MeV.
Thermal axion can be produce with photon electron scattering 
process {\bf II}, we will ignore this axion
unless the reheating temperature is very low 
(compared with the process {\bf IV}, it should be less than a few MeV).

To estimate the energy density of the axion KK mode for given $T$, 
multiply the entropy density at the temperature $s(T)$ 
and the mass of axion KK mode to the  yield.
We should count the number of KK modes for the allowed energy range,
$(E r_n)^N$ where $E$ is typically $T_R$. But for case {\bf III}, 
$T_R=\max\{m_\pi, T_R\}$.
$N$ is the dimension of the bulk where the produced particle exists.
$N=n$ and $N=m$ for case \{I, II\} and \{III, IV\} respectively.

\section{Calculation of the cosmological bounds}
In this section we 
describe the details of these calculations for two non-trivial cases,
$n=4, m=3$ at $M_*=1$ TeV and $n=3, m=2$ at $M_*=10$ TeV,
though we have calculated with computer 
all relevant cosmological bounds with allowed sets of
$m$ and $n$ and presented it in Fig.~3 and 4. 

This is the case with relatively low $T_R\leq {\cal O}(100)$ MeV, because of
large $f_{PQ}\sim  10^{14}$ GeV. 
Since the life-time of the axion KK mode is quite long for
$m_{A} < 100$ MeV and CMBR is stronger bound than 
the light element breaking bound, 
the bound from light element breaking is not relevant in these cases.

\subsection{Big Bang Nucleosynthesis Bound}
The axion KK mode with mass below $100$ MeV cannot decay before $1$
sec and will be restricted by BBN bound on neutrino species.
Let's assume that $T_R >10$MeV then
\begin{eqnarray}
\left.\frac{\rho_A}{s}\right|_{BBN} &\simeq &
D\int^{m_1}_{m_0} dm_A \ (m_A r_n)^m Y_A \nonumber\\
&\simeq&
m_1 Y_3 (m_1 r_n)^m 
\sim 2\cdot 10^{-3} \frac{M_{pl} T_R^3 m_1^{m+1}}{M_*^{m+2} f_\pi^2}
A_3  < 0.1 \mbox{ MeV}
\end{eqnarray}
where $m_1 = \max\{m_\pi, T_R\}$ and $D$ is normalization constant defined as
\begin{eqnarray*}
D\int^{E_2}_{E_1}  dm_A \times m_A^{m-1} r_n^m \equiv \#\mbox{ of KK modes 
between } E_1 \mbox{ and } E_2  .
\end{eqnarray*}
We used the torus compactification with uniform distance approximation
$V_n= r_n^n, V_m = r_n^m$ so that $D=m$.
This gives a bound for $n=4, m=3$ at 1 TeV
\begin{equation}
T_R < 90\, A_3^{-1/3} \mbox{ MeV} \simeq 80 \mbox{ MeV}
\end{equation}
and for $n=3, m=2$ at 10 TeV
\begin{equation}
T_R < 100\, A_3^{-1/3} \mbox{ MeV} \simeq 90 \mbox{ MeV}.
\end{equation}

\subsection{Over-closure Bound}
We can divide this bound by KK mass $m_2$.
If $m_{A}> m_2$, its life-time is shorter than the age of universe,
otherwise it will remains as a cold dark matter.
For $T_R <10$ MeV, the over-closure
is not a problem in the region we are interested. Therefore
$Y_3$ is most dominant source of axion KK mode in this case.
The total cold dark matter density  will be 
\begin{equation}
\frac{\rho_A}{s_0}\simeq D\int^{m_2}_{m_0} dm_A\ (m_A r_n)^m Y_3
< 3\cdot 10^{-6} h^2\, {\rm MeV}
\label{eq34a}
\end{equation}
where $m_2 \simeq 10$ MeV, $m_0 = r_n^{-1} \simeq 30$ keV
 for $n=4, m=3$ at 1 TeV and
$m_2 \simeq 1$ MeV, $m_0 \simeq 5$ keV for  $n=3, m=2$ at 10 TeV.
Using Eq.(\ref{Y3}), Eq.(\ref{eq34a}) becomes
\begin{equation}
2\cdot 10^{-3} m_2^{m+1} \frac{M_{pl} T_R^3}{f_\pi^2 M_*^{m+2}}A_3
< 3 \cdot 10^{-6} h^2 \,{\rm MeV}.
\end{equation}
We can get the bound on $T_R$ for $n=4, m=3$ at 1 TeV,
\begin{equation}
T_R < 70 \left(\frac{h}{0.7}
\right)^{2/3} A_3^{-1/3} \mbox{ MeV} \label{bd34a}
\end{equation}
and for $n=3, m=2$ at 10 TeV
\begin{equation}
T_R < 350 \left(\frac{h}{0.7}
\right)^{2/3} A_3^{-1/3} \mbox{ MeV} \label{bd23a}.
\end{equation}

We can consider another situation that the axion KK mode which can decay
into some relativistic dark matter $X$ dominantly.
In this case,
\begin{equation}
\frac{\rho_A}{s_0}\simeq D\int^{m_1}_{m_2} dm_A\, (m_A r_n)^m Y_3  
\frac{T_0}{T(m_A)}
< 3\cdot 10^{-6} h^2\, {\rm MeV},
\label{eq34b}
\end{equation}
where $T_0 \sim 2\times 10^{-13}$ GeV is current temperature of universe
and 
\begin{equation}
T(m_A) \simeq  \sqrt{
\Gamma_{a_{KK}\rightarrow X} M_{pl} } \sim
2\cdot 10^{-4} C_{a\gamma} Br^{-1/2}\frac{m_A^{3/2}M_{pl}^{1/2}}{f_{PQ}}
\label{decaytemp}
\end{equation}
is the temperature when axion KK mode with mass $m_A$ decays,
where $Br$ is the branching ratio for 
\begin{equation}
Br=\frac{\Gamma_{a_{KK}\rightarrow 2\gamma}}{\Gamma_{a_{KK}\rightarrow X} 
+\Gamma_{a_{KK}\rightarrow 2\gamma}}\simeq \frac{\Gamma_{a_{KK}\rightarrow
2\gamma}}{\Gamma_{a_{KK}\rightarrow   X}},
\end{equation}
which is the case that majority of axion KK modes decays into
the invisible relativistic dark matter $X$.
Then Eq.(\ref{eq34b}) becomes
\begin{equation}
2\cdot 10^{-9} Br^{1/2}\frac{f_{PQ}}{C_{a\gamma}} 
\frac{M_{pl}^{1/2} T_R^3 m_1^{m-1/2} }{M_*^{m+2} f_\pi^2}
A_3 \, {\rm MeV}
< 3\cdot 10^{-6} h^2 \,{\rm MeV} .
\end{equation}
For the reasonable range of $Br$, we can set $m_1=m_\pi$.
For $n=4, m=3$ at 1 TeV,
\begin{equation}
T_R <  12 Br^{-1/6} \left(\frac{h}{0.7}
\right)^{2/3} C_{a\gamma}^{1/3}A_3^{-1/3} \mbox{ MeV} 
\simeq 30 \mbox{ MeV (for }Br=1)
\label{bd34b},
\end{equation}
and for $n=3, m=2$ at 10 TeV,
\begin{equation}
T_R <  28 Br^{-1/6} \left(\frac{h}{0.7}
\right)^{2/3} C_{a\gamma}^{1/3}A_3^{-1/3} \mbox{ MeV}
\simeq 40 \mbox{ MeV (for }Br=1)
 \label{bd23b}.
\end{equation}

\subsection{Cosmological Microwave Background Radiation}
CMBR bound is the most severe during the time period $ 10^6
<\tau_A <10^{12}$ sec. The bound
\begin{equation}
\frac{\Delta \rho_\gamma}{s} \leq 2.5 \times 10^{-5} T_D
\label{cmbr}
\end{equation}
is actually weaker than other constraints if the branching ratio $Br$ is
 large. In this case, the  reheating temperature should be relatively small 
$T_R < 10$ MeV.
But if $Br$ is very
small, it gives stronger bound than other cosmological constraints.
$T_D$
is the same as Eq.(\ref{decaytemp}). If we assume $T_R> 10$ MeV,
and set $\Delta \rho_\gamma \simeq Br\times \rho_A(T_D)$,
Eq.(\ref{cmbr}) becomes
\begin{equation}
2\cdot 10^{-3} Br\cdot m_A^{m+1} \frac{M_{pl} T_R^3}{M_*^{m+2} f_\pi^2}A_3
\leq 5 \cdot 10^{-9} C_{a\gamma} Br^{-1/2} \frac{M_{pl}^{1/2} m_A^{3/2}
}{f_{PQ} }
\end{equation}
If $T_R \leq m_\pi$, the maximal value of KK mode mass is around
$m_\pi$. This leads the bound,
for $n=4, m=3$ at 1 TeV,
\begin{equation}
T_R < 2\cdot  10^{-2} Br^{-1/2} A_3^{-1/3}
C_{a\gamma}^{1/3}\, \mbox{ MeV}
\end{equation}
for $n=3, m=2$ at 10 TeV,
\begin{equation}
T_R < 4 \cdot 10^{-2} Br^{-1/2} A_3^{-1/3}
C_{a\gamma}^{1/3}\, \mbox{ MeV}.
\end{equation}
To have $T_R > 100$ MeV, we need a very small branching ratio  $\sim 10^{-7}$.
(This bound is not valid if $T_R<10$ MeV.)

\subsection{Diffused Photon Background}
Since the life-time of axion KK mode  is longer than $10^{14}$ sec in allowed
region, we obtain
most strong bound on reheating temperature from diffused photon background.
Let's consider three cases,
\\ {\bf Case 1.} $T_R>10$ MeV : In this case, the
majority of KK mode will decay before present time.
The observed bound when $800$ keV $< E< 30$ MeV is,
\begin{equation}
\frac{d{\cal F}}{d\Omega} 
< 78 \left(\frac{E}{ 1 {\rm  keV}}\right)^{-1.4} \mbox{cm}^{-2}
\mbox{sr}^{-1}\mbox{sec}^{-1}
\end{equation}
corresponds for theoretical prediction 
\begin{equation}
\frac{d{\cal F}}{d\Omega} = \frac {n_{\gamma} c}{4\pi} 
\simeq Br\frac{Y_3s_0 c}{4\pi}(m_Ar_n)^m.
\end{equation}
Here $s_0c\simeq 9\cdot 10^{13}$cm$^{-2}$sec$^{-1}$.
This will give the inequality,
\begin{equation}
 Br\times Y_3(m_Ar_n)^m < 6\cdot 10^{-16} \left(\frac{E}{ {\rm  MeV}}\right)^{-1.4}
\label{bd34c}
\end{equation}
where $E\simeq m_{A}/(2(1+z))$
\begin{equation}
1+z \simeq 4 \times 10^{11}\left(\Omega_0 h^2\right)^{-1/3} \left(
\frac{\tau_D}{\rm sec} \right)^{-2/3},
\end{equation}
The time of axion KK mode decay into two photons is
\begin{equation}
\tau_D\simeq Br\cdot \Gamma_{a_{KK}\rightarrow 2\gamma}^{-1}\simeq
3.7\cdot 10^7 Br \frac{f_{PQ}^2}{m_A^3}.
\end{equation}
Thus the present energy of diffused photon (for $\Omega_0\simeq 1$) is
\begin{eqnarray}
E&\simeq& 10^{-12} \left(\frac{\tau_D}{\rm sec}
\right)^{2/3}\left(\frac{h}{0.7}\right)^{2/3} m_A \nonumber\\ 
&\simeq& 10^{-18}C_{a\gamma}^{-4/3} Br^{2/3} \left(\frac{f_{PQ}}{\rm
GeV}\right)^{4/3} \left(\frac{h}{0.7}\right)^{2/3}
\left(\frac{10\rm MeV}{m_A}\right) \mbox{ MeV}.
\end{eqnarray}
Inserting this result into (\ref{bd34c}) and setting $h=0.7$ leads
\begin{equation}
2\cdot 10^{-3} Br \frac{M_{pl} m_A^m T_R^3}{f_\pi^2 M_*^{m+2}} A_3 <
10^{10} \left(\frac{f_{PQ}}{\rm GeV}\right)^{-1.9} C_{a\gamma}^{1.9} Br^{-0.9}\left(\frac{m_A}{10\rm
MeV}\right)^{1.4}
\label{ap17}
\end{equation}
and it gives a bound for $n=4, m=3$ at 1 TeV,
\begin{equation}
T_R< 2\cdot 10^{-2} \left(\frac{Br}{C_{a\gamma}}\right)^{-0.63} \left(
\frac{m_A}{10\rm MeV}\right)^{-0.53}A_3^{-1/3} \mbox{ MeV}
\end{equation}
and  for $n=3, m=2$ at 10 TeV,
\begin{equation}
T_R< 3\cdot 10^{-2} \left(\frac{Br}{C_{a\gamma}}\right)^{-0.63} \left(
\frac{m_A}{10\rm MeV}\right)^{-0.2}A_3^{-1/3} \mbox{ MeV}.
\end{equation}

\noindent{\bf Case 2.} $T_R <10 $ MeV, 
$Br < \Gamma_{a_{KK}\rightarrow2\gamma}/H_0$: \\
 In this case $Y\simeq Y_4$ and  axion KK mode will decay before 
present time.
$H_0 \simeq 2\cdot 10^{-42} h$ GeV is Hubble constant of present universe.
Then the relation with energy and life time become same as Eq.(\ref{ap17}),
\begin{eqnarray}
 Br\cdot Y_4\cdot(m_Ar_n)^m =&&
8 \cdot 10^{-10} Br \frac{M_{pl} m_A^m T_R}{ M_*^{m+2}} A_4 \nonumber\\
&&<
10^{10} \left(\frac{f_{PQ}}{\rm GeV}\right)^{-1.9} C_{a\gamma}^{1.9} Br^{-0.9}\left(\frac{m_A}{10\rm
MeV}\right)^{1.4}
\label{bd34e}
\end{eqnarray}
and it gives a bound for $n=4, m=3$ at 1 TeV
\begin{equation}
T_R<  0.06\left(\frac{Br}{C_{a\gamma}}\right)^{-1.9} \left(
\frac{m_A}{\rm MeV}\right)^{-1.6} A_4^{-1} \mbox{ MeV}
\end{equation}
and in $m_A \sim T_R$ limit,
\begin{equation}
T_R< 0.3\, Br^{-0.73} \mbox{ MeV}.
\end{equation}
For  $n=3, m=2$ at 10 TeV
\begin{equation}
T_R<  0.03\left(\frac{Br}{C_{a\gamma}}\right)^{-1.9} \left(
\frac{m_A}{\rm MeV}\right)^{-0.6} A_4^{-1} \mbox{ MeV},
\end{equation}
and in $m_A \sim T_R$ limit,
\begin{equation}
T_R< 0.1\, Br^{-1.2} \mbox{ MeV}.
\label{Eq31}
\end{equation}

\noindent{\bf Case 3.} $T_R <10 $ MeV, 
$Br > \Gamma_{a_{KK}\rightarrow2\gamma}/H_0$: \\
In this case,
 axion KK modes will not decay before the present time. 
Then the theoretical prediction becomes (see references, for
instance, Sec. 5.5 in \cite{Kolb-Turner})
\begin{eqnarray}
\frac{d{\cal F}}{d\Omega} &\simeq& Br \times\frac {Y_4s_0 c}{4\pi} 
\frac{\Gamma_{a_{KK}\rightarrow
2\gamma} }{Br H_0} \left( \frac{2E}{m_A}\right)^{3/2} (m_Ar_n)^m
\end{eqnarray}
and this will lead
\begin{equation}
2\cdot 
10^{-17}C_{a\gamma}^2\frac{M_{pl} T_R}{M_*^{m+2} H_0} \frac{m_A^{m+3/2}E^{3/2}
}{f_{PQ}^2} A_4 
< 6\cdot 10^{-16} \left(\frac{E}{ {\rm  MeV}}\right)^{-1.4} .
\label{bd34d}
\end{equation}
Approximately, for $n=4, m=3$ at 1 TeV,
\begin{equation}
T_R  \left(\frac{m_A}{ {\rm  MeV}}\right)^{9/2} < 2\cdot 10^6\left(
\frac{h}{0.7}\right)
C_{a\gamma}^{-2}A_4^{-1}\left(
\frac{E}{{\rm  MeV}}\right)^{-2.9} \mbox{ MeV}.
\end{equation}
We can approximate $m_A\sim T_R\sim E$, then $A_4 \sim C_{a\gamma}^2$ and
\begin{equation}
T_R < 5.5 \left(
\frac{h}{0.7}\right)^{0.12} C_{a\gamma}^{-0.48} \mbox{ MeV}.
\end{equation}
The case $n=3, m=2$ at 10 TeV, the majority of
thermal axion KK mode with $T_R>1$ MeV will 
decay before present time.  The condition 
$Br \simeq \Gamma_{a_{KK}\rightarrow2\gamma}/H_0$
and Eq.(\ref{Eq31}) 
will determine the reheating temperature
\begin{equation}
T_R < 2 \sim 3 \mbox{ MeV}.
\end{equation}
We did full calculations for all relevant sets of $m$ and $n$ with
computer and get consistent results.
For instance, you can see 
three different regions of DPBR bound in Fig. 1 and 2.

\begin{table}
\caption{PQ scales and lifetimes of axion and graviton KK modes
for $M_*=1$ TeV, where $M_{100}\equiv
\frac{m_{A}}{100\rm MeV}$}
\label{table1}
\begin{tabular}{ccccc}
$(m, n)$ &$r_n^{-1}[\mbox{MeV}]$& $f_{PQ}[\mbox{GeV}]$  
& $\tau_g[\mbox{sec}]$ & $\tau_a[\mbox{sec}]$   \\
\tableline
$(1,2)$ &$4\times 10^{-10}$ &  $ 5\times 10^{10}$  & 
$2\times 10^{7}M_{100}^{-4}$ &
$6\times 10^{7}M_{100}^{-3}$\\
$(2,3)$ &$6\times 10^{-5}$ &  $2\times 10^{13}$       & 
$1\times 10^{6}M_{100}^{-5}$ &
$8\times 10^{12}M_{100}^{-3}$\\
$(2,4)$ &$2\times 10^{-2}$ &  $ 5\times 10^{10}$  & 
$2\times 10^{11}M_{100}^{-5}$ &
$6\times 10^{7}M_{100}^{-3}$\\
$(3,4)$ & &  $ 3\times 10^{14}$ & 
$3\times 10^{7}M_{100}^{-6}$ &
$3\times 10^{15}M_{100}^{-3}$\\
$(2,5)$ &$0.7$ &  $1\times 10^{9}$          & 
$2\times 10^{14}M_{100}^{-5}$ &
$5\times 10^{4}M_{100}^{-3}$\\
$(3,5)$ & &  $2\times 10^{12}$       & 
$1\times 10^{12}M_{100}^{-6}$ &
$7\times 10^{10}M_{100}^{-3}$\\
$(4,5)$ & &  $2\times 10^{15}$       & 
$9\times 10^{9}M_{100}^{-7}$ &
$1\times 10^{17}M_{100}^{-3}$\\
$(3,6)$ &$7$ &  $ 5\times 10^{10}$  & 
$2\times 10^{15}M_{100}^{-6}$ &
$6\times 10^{7}M_{100}^{-3}$\\
$(4,6)$ & &  $2\times 10^{13}$       & 
$1\times 10^{14}M_{100}^{-7}$ &
$8\times 10^{12}M_{100}^{-3}$\\
%\hline
\end{tabular}
\end{table}

\begin{table}
\caption{PQ scales and lifetimes of axion and graviton KK modes
for $M_*=10$ TeV.}
\label{table2}
\begin{tabular}{ccccc}
$(m, n)$ &$r_n^{-1}[\mbox{MeV}]$ & $f_{PQ}[\mbox{GeV}]$  & $\tau_g[\mbox{sec}]$ & $\tau_a[\mbox{sec}]$   \\
\tableline
$(1,2)$ &$4\times 10^{-8}$ &  $2\times 10^{11}$  & 
$2\times 10^{9}M_{100}^{-4}$ &
$6\times 10^{8}M_{100}^{-3}$\\
$(2,3)$ &$3\times 10^{-3}$ &  $4\times 10^{13}$       & 
$3\times 10^{9}M_{100}^{-5}$ &
$4\times 10^{13}M_{100}^{-3}$\\
$(2,4)$ &$0.6$ &  $2\times 10^{11}$  & 
$2\times 10^{14}M_{100}^{-5}$ &
$6\times 10^{8}M_{100}^{-3}$\\
$(3,4)$ & &  $6\times 10^{14}$ & 
$1\times 10^{12}M_{100}^{-6}$ &
$9\times 10^{15}M_{100}^{-3}$\\
$(2,5)$ &$20$ &  $6\times 10^9$          & 
$1\times 10^{17}M_{100}^{-5}$ &
$8\times 10^{5}M_{100}^{-3}$\\
$(3,5)$ & &  $ 4\times 10^{12}$       & 
$2\times 10^{16}M_{100}^{-6}$ &
$5\times 10^{11}M_{100}^{-3}$\\
$(4,5)$ & &  $3\times 10^{15}$       & 
$4\times 10^{15}M_{100}^{-7}$ &
$3\times 10^{17}M_{100}^{-3}$\\
$(3,6)$ &$160$ &  $2\times 10^{11}$  & 
$2\times 10^{19}M_{100}^{-6}$ &
$6\times 10^{8}M_{100}^{-3}$\\
$(4,6)$ & &  $4\times 10^{13}$       & 
$3\times 10^{19}M_{100}^{-7}$ &
$4\times 10^{13}M_{100}^{-3}$\\
%\hline
\end{tabular}
\end{table}

\begin{figure}
\begin{center}
\epsfxsize 13cm
\epsfbox{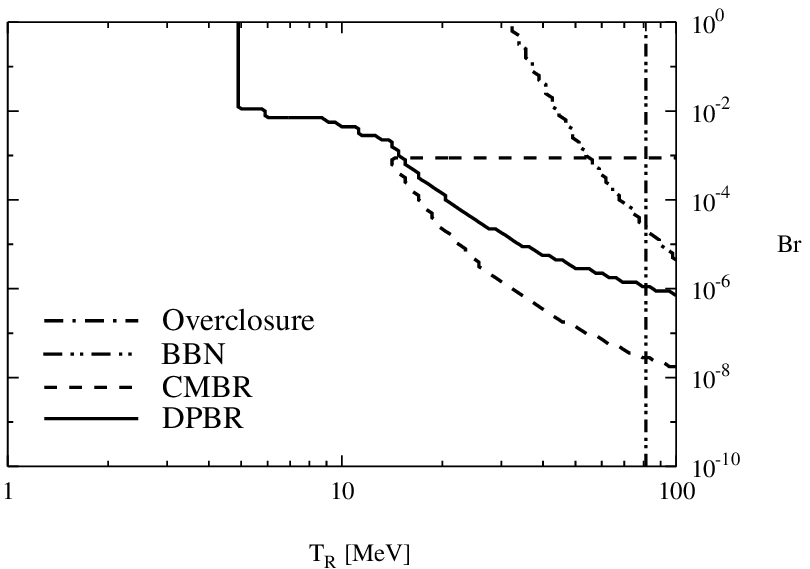}
\end{center}
%\vspace{-5mm}
\caption{The bound of $T_R$ and $Br$ for $M_*=1$ TeV and $n=4, m=3$,
The upper and right side of each line is excluded region.}

\vspace{5mm}
\begin{center}
\epsfxsize 13cm
\epsfbox{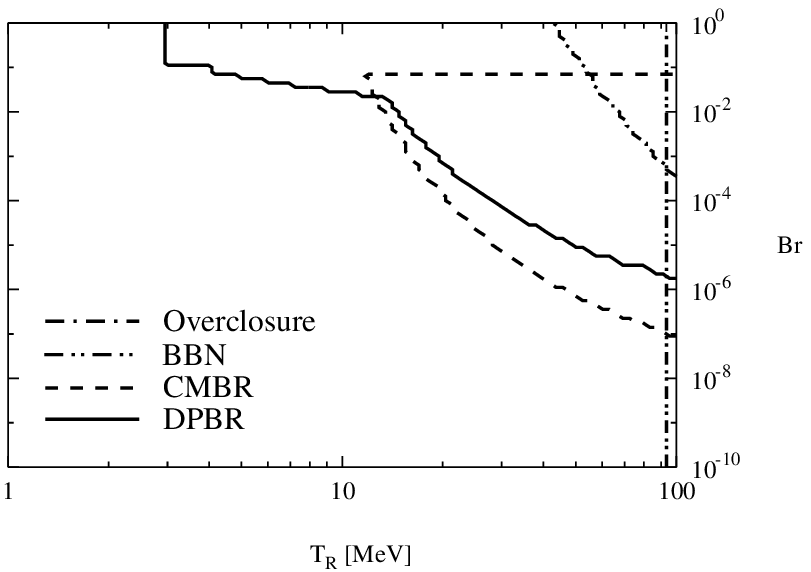}
\end{center}
\caption{The bound of $T_R$ and $Br$ for $M_*=10$ TeV and $n=3, m=2$. The
excluded region is same as Fig.~1.}

\begin{center}
\epsfxsize 13cm
\epsfbox{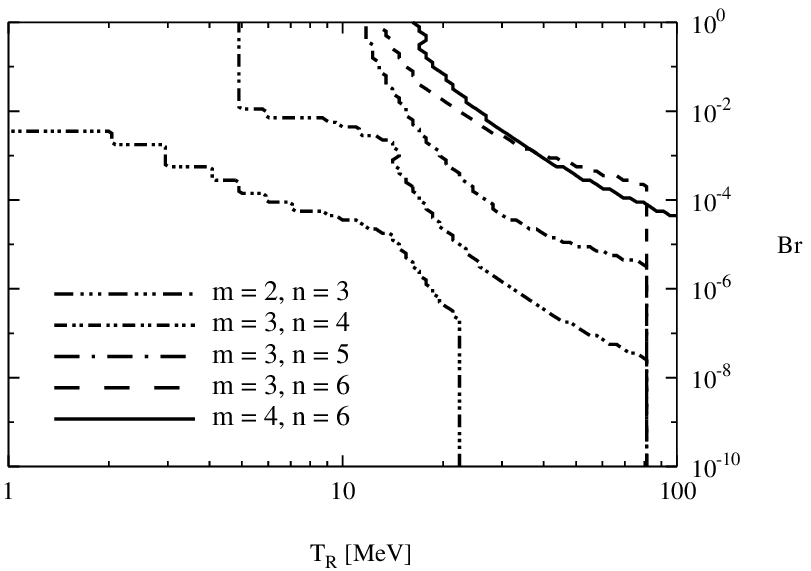}
\end{center}
\caption{The bound of $T_R$ and $Br$ for $M_*=1$ TeV. The upper and right side of
each line is excluded region.}

\begin{center}
\epsfxsize 13cm
\epsfbox{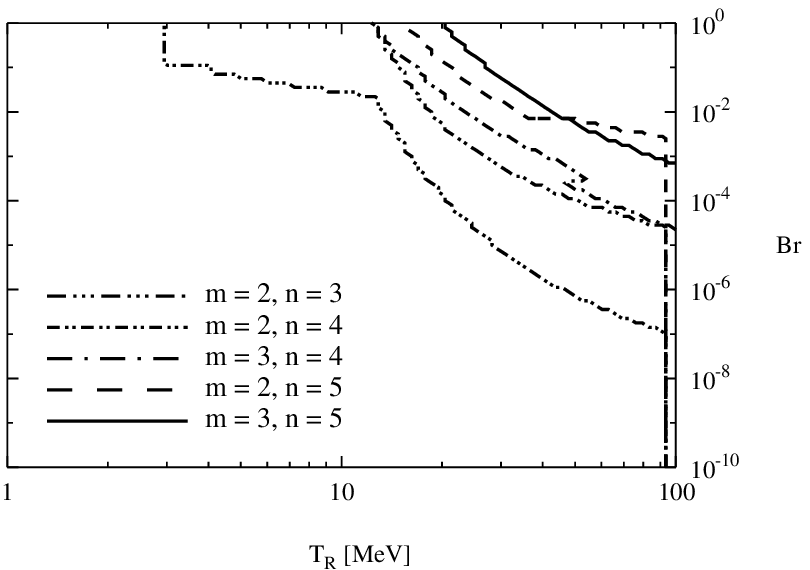}
\end{center}
\caption{The bound of $T_R$ and $Br$ for $M_*=10$ TeV. The
excluded region is same as Fig.~3.}

\begin{center}
\epsfxsize 13cm
\epsfbox{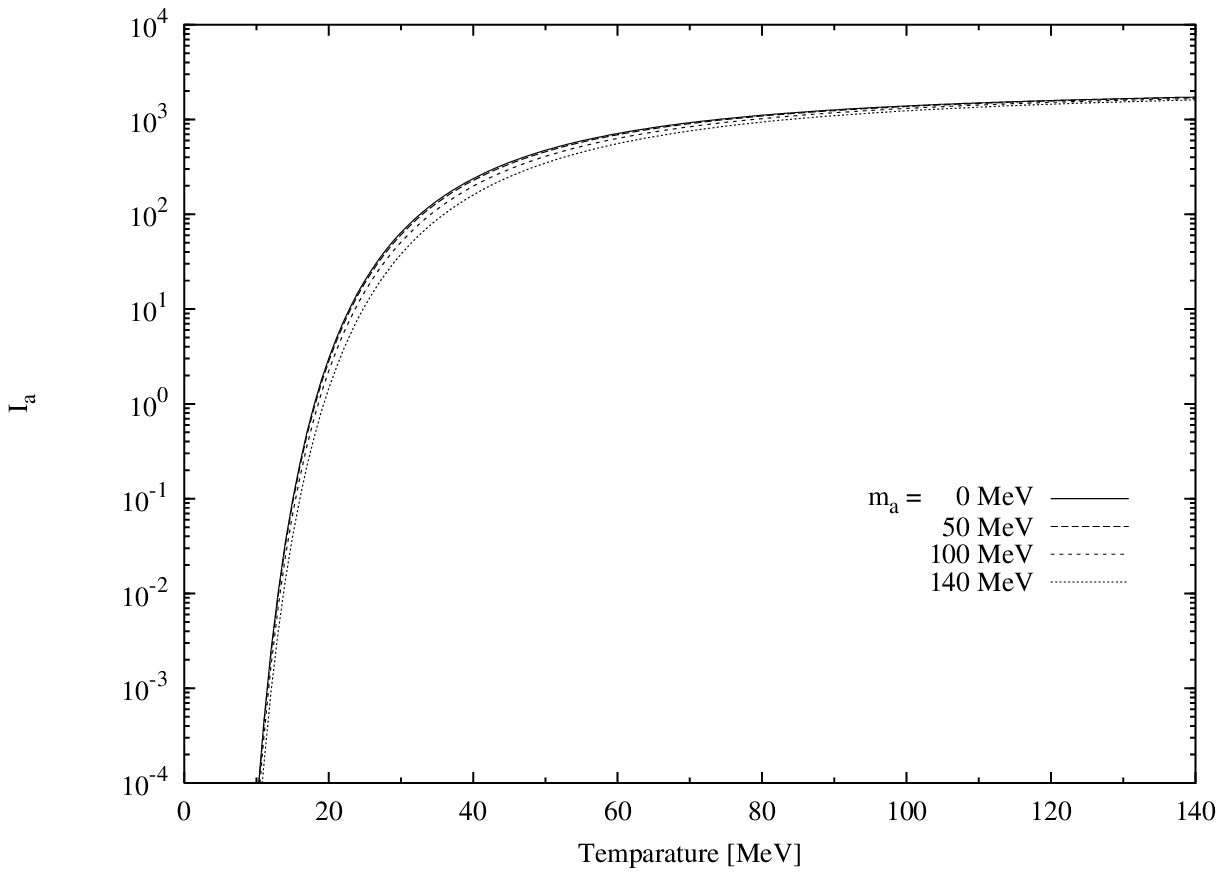}
\end{center}
\caption{The function $I(T)$ for various $m_{A}$ in the KK mode yield III.}
\end{figure}
\end{document}